Tight-Binding Theory of Lanthanum Strontium Manganate

Walter A. Harrison
Applied Physics Department
Stanford University
Stanford, CA 94305

An earlier analysis of manganese oxides in various charge states indicated that free-atom term values and universal coupling gave a reasonable account of the cohesion. This approach is here extended to $La_xSr_{1-x}MnO_3$ in a perovskite structure, and a wide range of properties, with comparable success. The cohesion, as a function of $x$, is rather well given. Magnetic and electronic properties are then treated in terms of the same parameters and the cluster orbitals used for cohesion. This includes an estimate of the Néel and Curie-Weiss temperatures for $SrMnO_3$, which is found to be an antiferromagnetic insulator. We estimate the magnitude of a Jahn-Teller distortion in $LaMnO_3$ which makes it also insulating. We find a magnetic state of $LaMnO_3$ with (100) ferromagnetic planes (due to a novel double-exchange for the distorted state), antiferromagnetically stacked, as observed. We estimate the corresponding Néel temperature and its volume dependence, and the ferromagnetic Curie-Weiss temperature which applies between the Néel and Jahn-Teller temperatures. We expect hopping conductivity when there is doping ($0<x<1$) and estimate it in the context of small-polaron theory. It is in accord with experiment between the Néel and Jahn-Teller temperatures, but below the Néel temperature the conduction appears to be band-like, for which we estimate a hole mass as enhanced in *large*-polaron theory. We see that above the Jahn-Teller temperature $LaMnO_3$ should be metallic as observed, and paramagnetic with a ferromagnetic Curie-Weiss constant which we estimate. Many of these predictions are not so accurate, but are sufficiently close to provide a clear understanding of all of these properties in terms of a simple theory and parameters known at the outset. We provide also these parameters for Fe, Co, and Ca so that formulae for the properties can readily be evaluated for similar systems.

## I. INTRODUCTION

We seek a sufficiently simple representation of the electronic structure of transition-metal compounds to allow us to estimate the entire range of properties by hand.[1-3] It would also provide a tool for interpreting experimentally observed properties. We do not attempt the high accuracy which highly computational techniques seem to promise[4]. Indeed there has been a full LDA+$U$ band calculation by Park[5] for $LaSrMnO_4$, a layered compound very similar to the alloy $La_xSr_{1-x}MnO_3$. It was able to address some of the questions we consider here, the stability of the antiferromagnetic state and the Jahn-Teller distortion, but the many-electron enhancement of gaps and polaron-like effects were beyond its reach. It was largely orthogonal to the present study. We are careful to include in a tight-binding context the same physics described by those techniques and do not rely on empirical parameters which might allow us to fit observed properties with an incorrect theory. This approach allowed us to study Heisenberg exchange in the 3$d$ monoxides[3], and learn that although superexchange dominated the second-neighbor interactions, an unanticipated direct exchange dominated nearest neighbors. These parameters gave a good account of the Curie-Weiss and Néel temperatures, their volume dependences, and related properties. The same approach was applied to the cohesive energy of the oxides of manganese[2] involving three different formal charge states, providing a clear picture of the origin of that cohesion, and its variation with charge state. An important finding, that the shift in $d$-state energy with change in charge state was small enough to be neglected for the cohesion, greatly simplified that picture. In particular, it meant that self-interaction corrections were small so that total energy changes could be calculated directly as the sum of changes in energy of occupied one-electron states. We seek now to extend this description to some more interesting perovskite compounds, based also upon manganese, where we may expect the same to be true. [We find, incidentally, that such self-consistent shifts are not small in all materials; we find them very large in the new superconductors, LaOFeP and LaOFeAs.]

Each manganese in the rock-salt-structure MnO is octahedrally surrounded by six oxygen ions and for the study of cohesive energy we described the electronic structure in terms of cluster orbitals centered on the Mn ions, sharing the oxygen orbitals with other Mn ions. For the tight-binding description of each cluster we used free-atom term values[6] from Ref. 1, showing separately that a self-consistent calculation gave sufficiently small changes in charge distributions that shifts in these values were not important for the cohesion. We used universal couplings between manganese $d$ states and oxygen $p$ states, such as[1]

$$V_{pd\sigma} = -(3\sqrt{15/2\pi})\hbar^2(r_d^3 r_p)^{1/2}/md^4. \tag{1}$$

La$_x$Sr$_{1-x}$MnO$_3$ has the same oxygen octahedron around each Mn and the same approach may be taken, and in fact the coupling from Eq. (1) differs only in the use of a $d$=2.01 Å, rather than the 2.22 Å for MnO in Ref. 2, increasing $V_{pd\sigma}$ to −1.614 eV and again $V_{pd\pi}$ = −$V_{pd\sigma}/\sqrt{3}$.

The free-atom onsite energies which enter are listed in Table 1, and were taken from Ref. 1 (except that La values were not given there, but are taken from the same Hartree-Fock tables[6]). We have included parameters for Fe and Co also, to facilitate treatment of other systems.  Also, Ca could be included at $\varepsilon_s = -5.32$ eV and $U_{ss} = 7.3$ eV.  The $\varepsilon_{d,}$ values had been calculated for equal occupation of up and down spins, and were corrected for full majority-spin shells, using the $U_x$ exchange from Ref. 1, as in Ref. 3 but here including Co.  These are good approximations to the removal energy of the corresponding electron from the neutral atom, except for $\varepsilon_d^{min}$ for Mn which has no minority-spin electrons; it is the energy to which a majority electron would go if its spin were flipped.   We found from self-consistent calculations in Ref. 2 for changing the formal valence from $Mn^{2+}$ (in MnO) to $Mn^{3+}$ (in $Mn_2O_3$) the removal energy of a second occupied $e_g$ level was lowered by only 0.44 eV if the first $e_g$ electron is removed. This was not enough to be important for the cohesion, so we could do the simplest theory with the same $\varepsilon_d^{maj}$ for all compounds (including $MnO_2$). We have redone that calculation for the smaller spacing in $LaMnO_3$ and found that the lowering of the second removal energy due to removing the first is very similar at 0.66 eV, and we neglect it also here.

Table 1. Tight-binding parameters in eV, except for $r_d$.

| | La | Sr | Mn | Fe | Co |
|---|---|---|---|---|---|
| $\varepsilon_s$ | −5.34 | −4.86 | −6.84 | −7.08 | −7.31 |
| $\varepsilon_d^{maj}$ | −6.80 | − | −17.22 | −18.06 | −19.30 |
| $\varepsilon_d^{min}$ | − | − | −14.10 | −15.78 | −16.24 |
| $\varepsilon_d^{min*}$ | − | − | − 8.50 | − 9.88 | − 9.94 |
| $U_d$ | − | − | 5.6 | 5.9 | 6.3 |
| $U_{ss}$ | 7.3 | 6.9 | 8.2 | 8.3 | 8.5 |
| $U_{dd}$ | − | − | 16.00 | 16.65 | 17.28 |
| $r_d$ | − | − | 0.925Å | 0.864Å | 0.814Å |

   $\varepsilon_d^{min}*$ is the energy to be associated with an empty minority state,  which is coupled to the occupied oxygen $p$-like states.  It would be $\varepsilon_d^{min}+U_{dd}$ for an isolated neutral atom, where $U_{dd}$ is the increase in the removal energy of a $d$ electron in the neutral atom if another electron is brought from large distance and placed in another $d$ state.  In the solid we use a screened value, $U_d$, equal to the change in energy of a $d$ electron if an $s$ electron (rather than a distant electron) in the atom is transferred to the $d$ shell, as in Refs. 1, 2 and 3.

$$\varepsilon_d^{min}* = \varepsilon_d^{min} + U_d. \tag{2}$$

The appropriateness of this screened electron affinity for minority-spin levels seems quite clear.  Our use of $\varepsilon_d^{maj}$ , without the $U_d$ , for calculating the  empty upper $edmaj$ level in Fig. 1 is less clearly appropriate.  It was motivated by our finding that there were only small changes in the term values with change in charge state, and use of $\varepsilon_d^{maj}$ is consistent with that.   We would expect to do the same with one minority-spin level for $Fe^{3+}$ since that level would be occupied in the free atom, and this may be even less clearly appropriate.   $U_d$ was taken from Ref. 1 and is listed in Table 1.  So also are $U_{dd}$ and $U_{ss}$ ($U_{ss}$ is the downward shift in energy of an $s$ electron if the second $s$ electron is removed from the neutral atom. These were obtained in Ref. 1 from the experimental free-atom spectral tables[7] for Mn, where values were also obtained for Fe by scaling the values for Mn for each element by the square root of the free-atom term value; we did the same here for cobalt.  These values, plus the  $\varepsilon_p = -16.77$ eV, $r_p = 4.41$ Å, and $U_{pp} = 14.47$ eV for oxygen[1] are the only tight-binding parameters we will need in our analysis.

   For the $d$ state in each cluster orbital there is a linear combination of $p$ states on the neighboring oxygen ions of the same symmetry; e. g., for an $x^2-y^2$ $d$ state it is one half times the sum of sigma-oriented $p$ states on the neighbors in the $\pm x$ direction minus one half times the sum of sigma-oriented $p$ states in the $\pm y$ direction. The coupling between the $d$ state and the combination of $p$ states is obtained using the Slater-Koster Tables (e. g., Ref.1, p.546) as $V_2^{eg} = \sqrt{3}\ V_{pd\sigma} = -2.80$ eV. There are two such $e_g$ states (the other is of symmetry $3z^2-r^2$) with the same coupling.  There are also three $t_g$ states (of symmetry, $xy$, $yz$, and $zx$) for which the coupling is $V_2^{tg} = 2V_{pd\pi} = 1.86$ eV.  We write half the difference in energy as $V_3^{min} = (\varepsilon_d^{maj} - \varepsilon_p)/2 = -0.225$ eV for majority-spin states and find the majority-spin $e_g$ levels, measured from the oxygen $\varepsilon_p$, as[2]

$$\varepsilon_{eg}^{maj} = V_3^{maj} \pm \sqrt{(V_2^{eg2} + V_3^{maj2})}, \tag{3}$$

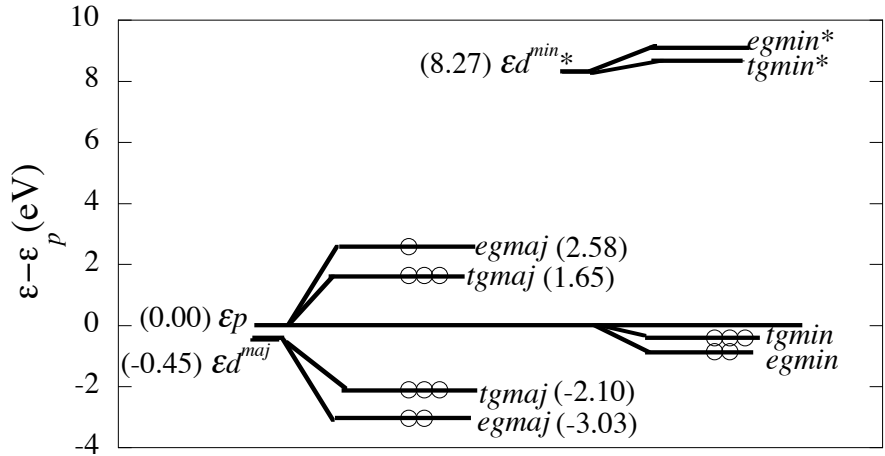

Fig. 1. The energy levels (in eV, measured from the oxygen *p*-state energy) for majority-spin (designated by *maj*) and minority-spin (designated by *min*) cluster orbitals for LaMnO₃. Doubly-degenerate states of $e_g$ symmetry are distinguished from triply-degenerate $t_g$ states. Circles indicated occupied states. The upper *egmaj* level would have an additional electron in MnO, and no electrons in SrMnO₃, but otherwise the diagrams are qualitatively the same.

For majority-spin $t_g$ levels $V_2^{eg2}$ is replaced by $V_2^{tg2} = 4V_{pd\pi}^2$. For minority-spin levels, $V_3^{maj}$ is replaced by $V_3^{min} = (\varepsilon_d^{min*} - \varepsilon_p)/2 = 4.14$ eV. The resulting levels are shown in Fig. 1 and form the basis for our study. We may proceed to the cohesive energy, beginning with SrMnO₃.

## II. COHESIVE ENERGY

In SrMnO₃ the Sr contributes its two *s* electrons to the oxygens, as well as does each Mn. We may proceed exactly as for the oxides[1,2] for the corresponding change in energy with the transfer of

$$\Delta E_{s\,to\,p} = 2(\varepsilon_p - \varepsilon_s(\text{Sr})) + 2(\varepsilon_p - \varepsilon_s(\text{Mn})) + U_{ss}(\text{Sr}) + U_{ss}(\text{Mn}) + 5U_{pp} + E_{es}. \tag{4}$$

[Only the last electron added to the three oxygens enters at $\varepsilon_p + 2U_{pp}$, the other three at $\varepsilon_p + U_{pp}$.] For LaMnO₃ the same formula applies with $\varepsilon_s(\text{Sr})$ replaced by $\varepsilon_s(\text{La})$ and $U_{ss}(\text{Sr})$ replaced by $U_{ss}(\text{La})$, raising the energy of transfer $\Delta E_{s\,to\,p}$ by 1.36 eV . $E_{es}$ is the electrostatic energy per molecule for a perovskite structure with charges +2$e$ on the Mn and Sr sites and $-^4/_3e$ on the oxygen sites. This is just $^4/_9$ of the energy for LaAlO$_3$ with +3$e$ on the La and Al sites and $-2e$ on the oxygen sites, which we calculated as $-22.35e^2/d$ in Ref. 8. $^4/_9$ths of that is $-71.16$ eV here, applying to both SrMnO$_3$ and LaMnO$_3$ since at this stage we have transferred just two electrons in both cases and we take the lattice spacing the same. The combination of both terms contributes $\Delta E_{s\ to\ p} = -27.39$ eV to the energy of formation of SrMnO$_3$ and $\Delta E_{s\ to\ p} = -26.03$ eV for LaMnO$_3$.

There are also contributions to the energy from the coupling $V_{pdm}$ which are calculated in terms of cluster orbitals as in Eq. (3). We see that the lowering in energy of the lower state due to the coupling was $-\sqrt{(V_2{}^2+V_3{}^2)}+V_3$ for each state. For both SrMnO$_3$ and LaMnO$_3$ only these lower-energy minority-spin cluster orbitals are occupied and we may add the five shifts to obtain a contribution to the cohesive energy of $\Delta E_{min} = -2.91$ eV, the same for both compounds. At this stage in the formation, all majority-spin states are occupied and the upward shift of the upper states cancels the downward shift of the lower states so there is no contribution to the cohesion. There is also coupling between the empty Mn $s$ states and the oxygen $p$ states, but its small effect was neglected in Refs. 1 and 2 and here.

For SrMnO$_3$ we must also transfer two majority-spin $d$ electrons, from $e_g$ states which are higher in energy than the $t_g$ states, to nonbonding oxygen states, as we did in MnO$_2$. The energy of these upper $e_g$ levels was obtained with the plus sign in Eq. (3) and shown in Fig. 1 as 2.58 eV above the nonbonding $p$ states, for a total contribution to $\Delta E_{d\ to\ p}$ of $-5.16$eV per Mn.

For LaMnO$_3$ we transfer one $d$ electron from the Mn, gaining 2.58 eV, but also one from the La. We estimate this energy gain as $\varepsilon_d$(La)$-\varepsilon_p$ =9.97 eV for a total contribution to $\Delta E_{d\ to\ p}$ of -12.55 eV per Mn. This is treating the La $d$ electron the same as the Mn $d$ electron, though neglecting coupling since the state is well removed in energy. The predicted heats of atomization (cohesive energies) are then both exothermic with magnitudes from the three terms of 35.46 eV per Mn for SrMnO$_3$ and 41.49 eV for LaMnO$_3$. We would expect it to vary linearly as a function of $x$ for La$_x$Sr$_{1-x}$MnO$_3$.

The heat of formation for this series of compounds has indeed been measured by Rømark, et al.[9], [ the heat of formation is for separation to the elements, but we add the energy to separate the elements to atoms from Kittel[10]. ] and *does* vary linearly with $x$, though with deviations at the end points due to nonstoichiometries. Extrapolating the linear portion for the heat of formation from the elements, and adding the atomization energy of the elements gives exothermic magnitudes of 25.31 eV for SrMnO$_3$ and 30.42 eV for LaMnO$_3$. We have again overestimated the total cohesion but the predicted difference of 6.03 eV for the two compounds is close to the experimental difference of 5.11 eV. That difference is dominated by the extra energy gain in transferring an electron from a La $d$ state, rather than an upper MnO cluster state.

An important point is that the one extra electron in an $e_g$ state for a La substituted in SrMnO$_3$ could be on *any* Mn ion, and in fact in either of the $e_g$ states, and could jump to a neighboring site, providing electrical conductivity to the system. As in semiconductors the added electron can dope the system to make it conducting. It is true that as a negative charge it will be attracted to the La (or Fe for that doping) which produced it, it having one more positive charge than the Sr, or Mn, it replaced but that was also true in semiconductor doping. We shall see that such partially-filled $e_g$ levels will have other important consequences for LaMnO$_3$.

The analysis to this point has been based upon the energy of clusters, treated as independent. The essential features of the electrical and magnetic properties depend directly upon the interaction *between* clusters. The experimental findings for La$_x$Sr$_{1-x}$MnO$_3$ have been reviewed and interpreted by Zhou and Goodenough[11], and we use their compilation of data.

## III. ELECTRICAL AND MAGNETIC PROPERTIES OF SRMNO$_3$

For $x = 0$, SrMnO$_3$, we have Mn$^{4+}$, with the upper majority $e_g$ states empty and all majority $t_g$ states occupied (see Fig. 1). We correctly expect this system to be insulating, and though the energy difference between the filled majority $t_g$ and the empty majority $e_g$ levels is only $\sqrt{(V_2^{eg2} + V_3^{maj2})} - \sqrt{(V_2^{tg2} + V_3^{maj2})} = 0.93$ eV, this corresponds to a band gap with a very large additional enhancement of $U_d = 5.6$ eV , so we still expect insulating behavior, to which we return in Section V. [We saw in Ref. 2 that the familiar Coulomb enhancement of band gaps beyond one-electron theory (e. g., Ref. 1) corresponds approximately in these systems to $U_d$.]

We expect antiferromagnetic spin aligments due to a Heisenberg exchange through the oxygen, such as was calculated in Ref. 3. For MnO there was a relatively large direct coupling of[3] $\Delta E_1 = 30$ meV ($\Delta E_1$ is the energy for parallel alignment of spins for two nearest-neighbor Mn ions minus that for antiparallel spins.) from direct overlap of nearest-neighbor Mn ions, and a smaller superexchange of $\Delta E_2 = 8$ meV for second neighbors which is more relevant here since in the perovskite structure we do not have the overlapping nearest neighbors, only nearest-neighors with an oxygen directly between. In Ref. 3 we obtained that superexchange contribution as [Ref. 3 , Eq. (3), with an erroneous factor of 2 before the $V_{pd\sigma}^4$ eliminated.] a $\Delta E_2$ equal to the energy of a pair of *second* neighbors if their spin was parallel, minus that if it was antiparallel, as

$$\Delta E_2 = 4 \frac{V_{pd\sigma}^4 + 2V_{pd\pi}^4}{(\varepsilon_d^{min*} - \varepsilon_p)^2(\varepsilon_d^{min*} - \varepsilon_d^{maj})}. \tag{5}$$

An important feature of this formula is the factor 4 in front, which would be only 2 in LDA theory, and is shown most directly to be correct in Appendix A of Ref. 3, but also in Refs. 1 and 2. This contribution to the energy comes directly from the interaction [in perturbation theory, a $V_{pdm}$ coupling the $d$ state to the oxygen $p$ state times a $V_{pdm}$ coupling the $p$ state to the other $d$ state, divided by the energy difference] between an occupied cluster orbital on one site and an empty cluster orbital of the same spin on a neighboring site. The first term in the numerator arises from the $e_g$ states and is 9/2 of the second term [$V_{pd\pi} = -V_{pd\sigma}/\sqrt{3}$] . It is eliminated for SrMnO$_3$ because the majority-spin $e_g$ states are empty, and even with spacing $d = 2.01$ Å, the remaining energy difference is quite small at $\Delta E_2 = 10.1$ meV. There is an additional contribution to the total energy due to occupying only the lower majority-spin $e_g$ states, which we included before, but it is independent of the relative spin orientations of neighbors.

In Ref. 3 we combined expressions from the literature relating $\Delta E_2$ with a Heisenberg exchange energy, $2J\mathbf{S}_i \cdot \mathbf{S}_j$, and the Néel temperature with the $J$. For the simple-cubic structure (appropriate for perovskites here, and equal to the second-neighbor contribution in the rock-salt structure) that gave

$$k_B T_N \approx (S+1)\Delta E/S, \tag{6}$$

and

$$k_B \theta \approx -(S+1)\Delta E/2S \tag{7}$$

for the Curie-Weiss constant $\theta$, and $1/k_B = 11.6$ $^oK$/meV. Note $S=5/2$ in MnO, $S=2$ in LaMnO$_3$, and $S=3/2$ in SrMnO$_3$.

These formulae were successful in Ref.3 and we use them here, though we recognize that our calculation of $\Delta E_2$ was a mean-field calculation with all spins up or all down on each ion. Use of that same mean-field approach for the magnetism would give $k_B T_N = 3\Delta E$ rather than $5/3\Delta E$. In any case, these all may be thought of as sums of contributions over the six neighbors, and the important differences, when $\Delta E$ varies between different neighbors, would be qualitatively the same. For SrMnO$_3$ the Néel temperature is given from Eq. (6) by $k_B T_N = {}^5/_3 \Delta E_2$, equal to 16.9 meV, or $T_N = 195^oK$, comparable to the observed[12] $260^oK$ for cubic SrMnO$_3$, agreement similar to the predictions in Ref. 3.

Above the Néel temperature the moments are disordered, but when a magnetic field is applied the alignment of each moment by the field is influenced by the partial alignment of its neighbors, leading to a Curie-Weiss behavior of the magnetic susceptibility, $\chi = C/(T-\theta)$. $\theta$ is frequently defined with the reverse sign in studies of antiferromagnetism, and we followed that format in Ref. 4. Here we choose it as the intercept with the $T$ axis for a Curie-Weiss plot of $1/\chi$ versus $T$, negative for antiferromagnets. Eq. (7) gives minus half $T_N$, or $-98^oK$, but we have not found an experimental value for SrMnO$_3$.

## IV. LAMNO$_3$ AND THE JAHN-TELLER DISTORTION

For $x=1$, LaMnO$_3$, we have Mn$^{3+}$ for every Mn ion, and either $e_g$ state could be occupied on each, so the question of conductivity is not so clear. However, this single occupation of a degenerate state is exactly the condition for a Jahn-Teller distortion, and Zhou and Goodenough[11] have indicated that this occurs, an elongation of $\varepsilon = 0.08$ in a $z$ direction and contraction of $-\varepsilon/2$ in the $x$ and $y$ directions. We may independently estimate the size of the effect by noting that the change in energy of the $x^2-y^2$ state linear in the distortion is $6V_{pd\sigma}\varepsilon/\sqrt{(V_3^{maj2}+3V_{pd\sigma}^2)} = 5.57\varepsilon$ eV. There will be an elastic term equal to $3(c_{11}-c_{12})\varepsilon^2/4$ with $c^* \equiv (c_{11}-c_{12})/2 = 47.8$ GP $= 0.298$ eV/Å$^3$ determined experimentally by Darling, et al.[13] for La$_{0.83}$Sr$_{0.17}$MnO$_3$ at 200$^oK$. Multiplying the energy per unit volume by the volume $(2d)^3$ per Mn gives an energy $12c^*d^3\varepsilon^2 = 29.1\varepsilon^2$ eV per Mn. We shall relate this to a spring-constant model later, but it is not necessary here. The energy is minimum at $\varepsilon=0.096$, close to the observed[11] 0.08, with a net gain of $-0.27$ eV per Mn. At the observed $\varepsilon=0.08$ the coupling to neighbors in the $z$ direction drops to $V_{pd\sigma}^z = -1.19$ eV and the coupling to neighbors in the $xy$ plane increases to $V_{pd\sigma}^{xy} = -1.88$ eV.  The upper $3z^2-r^2$ states are lower than the $x^2-y^2$ states and are occupied, with  only the upper $x^2-y^2$ states empty, again insulating, as observed.  [Our estimate also included a solution $\varepsilon=-0.096$, with the $3z^2-r^2$ empty instead, but we continue with the observed distortion.]  This estimated 0.27 eV is a large energy compared to that for the antiferromagnetic-to-normal-state change in $SrMnO_3$, so we expect a high transition temperature $T_{JT}$.  It *is* much higher at the observed[11] $T_{JT}=750^oK$, but we have not carried out the analysis, involving the difference in vibration spectra, for such a Jahn-Teller transition.

We look also at the magnetic properties of this Jahn-Teller insulating state, proceeding with the observed $\varepsilon = 0.08$.  Having here a partially emptied *majority* state is analogous to having in Ref. 3 the partially occupied *minority* states in FeO and CoO, and we now eliminate from Eq. (5) only the contribution of the $x^2-y^2$ state.  For neighbors in the $z$ direction with antiparallel spins we have the full contribution from $3z^2-r^2$ with the $V_{pd\sigma}{}^z$ coupling (13.4 meV), and the contribution (1.5 meV each) of the states $zx$ and $yz$ with $V_{pd\pi}{}^z=-V_{pd\sigma}{}^z/\sqrt{3}$.  There are no contributions for parallel spins (full states coupled to full, empty to empty) and we obtain $\Delta E^z = 16.4$ meV.

For neighbors in the $xy$ plane it is more complicated.  The $t_g$ contribution is larger than in the $z$ direction at 18.6 meV, because of the larger $V_{pd\pi}{}^{xy}$, with antiparallel spin, and zero with parallel spin.   The term in the numerator from the $3z^2-r^2$ is $1/16\ V_{pd\sigma}{}^{xy\ 4}$ (the coupling to each is $-V_{pd\sigma}{}^{xy}/2$, from the Slater-Koster Tables, e. g., Ref. 1, p. 546), a contribution of 5.2 meV.  There is no contribution from the empty $x^2-y^2$ state (empty for both minority and majority spin).  The difference now is that for *parallel* spin alignment there is now a contribution from the coupling of the occupied $3z^2-r^2$ majority-spin state to the empty $x^2-y^2$ majority-spin state. The corresponding terms in the numerator of Eq. (5) are $(\sqrt{3}V_{pd\sigma}/2)^2(-V_{pd\sigma}/2)^2$ (again from the Slater-Koster Tables), and for these $\varepsilon_d^{min}*=-8.50$ eV is everywhere replaced by $\varepsilon_d^{maj}*=\varepsilon_d^{maj}+U_d=-11.62$ eV.  This gives a very large energy shift of $-63.1$ meV for *parallel* spin, subtracting from the other terms to give a net difference of $-39.3$ meV, a very strong ferromagnetic interaction in the $xy$ planes.  This ferromagnetic contribution is related to what is called *double exchange*, generally described as a broadening of the levels into bands – if the spins are aligned – and the populating of only the lower states.   In $LaMnO_3$ with the Jahn-Teller distortion, this double exchange might better be described as adding the effect of coupling between the empty majority $x^2-y^2$ state and the full majority $3z^2-r^2$ state on a neighbor, which we just evaluated.

Thus we find a net ferromagnetic coupling within the (100) planes, with the antiferromagnetic superexchange between planes unaffected except by the reduced coupling from the distortion.  Zhou and Goodenough[11] indeed indicate that this is the magnetic structure of $LaMnO_3$ below the Néel temperature, or *Curie* temperature if one uses terminology for ferromagnetic systems.  We may estimate that Néel temperature. The energy of the ferromagnetic planes is the same as that of antiferromagnetic planes with $\Delta E^{xy}$ changed in sign, so we may use Eq. (6), with $S=2$, and an average $\Delta E$ of $(2\times39.3+16.4)/3=31.7$ meV, giving $T_N=551^oK$, compared to the observed[11] $140^oK$ .  In contrast for the Curie-Weiss constant, the double-exchange term contributes negatively to the constant and the superexchange terms positively, so the appropriate $\Delta E$ is $(-2\times39.3+16.4)/3=-20.7$ meV. Eq. (7) with $S=2$ gives $\theta=180^oK$. Zhou and Goodenough[11] give a value smaller than their $T_N$, as we expect, $\theta = 52K^o$, but Dyakonov, et al.[14] find a value equal to the Néel temperature, in their measurements $\theta = T_N = 153^oK$. All indicate net ferromagnetic coupling.

We may also note in passing that since the $V_{pd\pi}$ vary as $1/d^4$ (Eq. (1)) and $T_N$ varies as $\Delta E$ in Eq. (6), which varies as $V_{pd\pi}^4$ (Eq. (5)), we would predict that $T_N$ should vary with volume $\Omega$ as $\partial\log T_N/\partial\log\Omega = -16/3 = -5.3$. In fact Zhou and Goodenough[11] found a value of $-5.3$ for LaMnO$_3$, but they found values of $-3.8$ and $-3.0$ for CaMnO$_3$ and YCrO$_3$, respectively, where we would also predict $-5.3$. Further, they indicated complications from changes in the O-Mn-O angle with pressure.

## V. ELECTRICAL PROPERTIES, POLARONS SMALL AND LARGE

We have seen that both SrMnO$_3$, and LaMnO$_3$ below the Jahn-Teller temperature, have the occupied states separated in energy from the empty states, and both were antiferromagnetic. Whether they are described by local cluster states, or full and empty bands, does not matter for electrical conduction, unless the system is doped, as by substituting Sr atoms for La atoms. Then the resulting majority-spin holes of $3z^2-r^2$ symmetry may move from cluster to cluster, but particularly so in the $xy$ plane where the neighboring $3z^2-r^2$ majority-spin states – which have the same energy – have the same spin and are therefore coupled. We might expect partly filled energy bands to be produced, and a double exchange for such bands, mentioned in the last section, to produce the observed[11] ferromagnetic metallic state.

It is quite simple to write down the tight-binding $pd$ bands, as in Eq. (3) of Ref. 2, for this two-dimensional motion, or to study the bands from an LDA calculation[5]. However, Zhou and Goodenough[11] have suggested small-polaron behavior in this regime between the Néel temperature and the Jahn-Teller temperature, and we also might expect that the distortions of the lattice by the carriers, analogous to the Jahn-Teller distortions, would be large enough to self-trap the carriers and allow only hopping conductivity, describable as small-polarons. Then the bands would have limited meaning and we should use small-polaron theory for the conduction.

We may check on the small-polaron supposition by Zhou and Goodenough by estimating the polaron energy relative to the phonon energy. We note that a hole in the cluster simply nullifies – for that cluster – the linear term we had with only the $3z^2-r^3$ orbital occupied; the shift in energy (now relative to the Jahn-Teller state) of the system linear in distortion has the same magnitude as the shift we found in the preceding section, $5.57\varepsilon$ eV. In our calculation of the Jahn-Teller distortion we included the elastic energy per manganese ion under the shear distortion for the *entire* lattice. When we look at distortion in a single cluster, the elastic energy will be approximately twice the value we obtained there, now $58.2\varepsilon^2$ eV. [If we think of springs between Mn-O neighbors, the energy change per Mn was only from its six neighboring springs when we deformed the entire lattice, but now the neighboring oxygens stretch also the springs to their stationary neighbors.] Then the equilibrium shift is half as large, $\varepsilon_0 = 0.04$, giving a polaron – or relaxation – energy also half as large at $W_p = 0.135$ eV.

We may also write out the kinetic energy $3M_O d^2 (d\varepsilon/dt)^2/2$ (note that the lateral oxygens shift only by $\varepsilon d/2$) with $M_O$ the mass of an oxygen ion. [We now take $d^{xy} = d =$ 2.01 Å for the cubic structure for simplicity.] The canonical momentum is $P = 3M_O d\, d\varepsilon/dt$, and with the potential energy $24c*d^3\varepsilon^2$ just given, we find ( e. g., Ref. 15) a harmonic oscillator frequency $\omega_0 = 1.10 \times 10^{14}$ rad./sec. corresponding to $\hbar\omega_0 = 0.0723$ eV, about half $W_p$. Such polaron energies twice the $\hbar\omega_0$ are indeed consistent with small-polaron behavior.

The classic reference for small polarons is by Emin and Holstein[16], a very intricate analysis for a single state on each atom in a triangular lattice. It is very difficult to follow just what approximations are made, but we can see from a much cruder derivation of their result what they are. They think in terms of a single vibrational mode of frequency $\omega_0$ (radians per second), a longitudinal optical mode. They imagine a relaxation of the lattice if a carrier is present in a particular site, lowering the energy by $W_p$. Then for the carrier to jump to the next site, which initially has no relaxation, requires overcoming a barrier which they give as $W_p/2$. This is not obvious, but is achieved by the site at which the electron sits relaxing half the equilibrium value, costing ¼ $W_p$, and the target site also relaxing half the equilibrium value in the opposite direction, for the other ¼ $W_p$. [The author is indebted to David Emin for explaining this, and other, points to him.] If the attempt frequency is $\omega_0/2\pi$, the rate at which transfer would occur would be $(\omega_0/2\pi)\exp(-W_p/2k_BT)$. This prefactor is derived by Emin and Holstein, not assumed. If a field $E$ is applied to the right, the barrier to the right will be lowered to $W_p/2 - eEa$, if the neighboring site is $a = 2d$ to the right, and the same increase will add to the barrier to the left. Inserting those corrections, expanding to first order in $E$, and multiplying by the distance traversed gives the net velocity of the carrier, which multiplied by the carrier density $N$ and charge $e$ gives the current density

$$j = N (4e^2Ea^2/W_p) (\omega_0/2\pi) (W_p/2k_BT) \exp(-W_p/2k_BT). \qquad (8)$$

The final two factors reach a peak of $\exp(-1)$ at $k_BT = W_p/2$, or $T = 783^\circ$K. This is close to their result but they had the coupling (which they called $J$) subtracted from $W_p/2$ in the exponent, presumably to be identified with the shift in the neighboring levels from this coupling. Their prefactor also differed by a factor ¾ from the triangular-lattice geometry, rather than the square geometry we used here. In the case of the perovskites, we have a shift from a cluster shear of $\underline{e_g}$ symmetry (a *transverse* optical mode).

Substituting the $W_p$ and $\omega_0$ obtained above into Eq. (8) gives a conductivity

$$\sigma = j/E = 2067 \times x(783^\circ K/T)\exp(-783^\circ K/T) \text{ (ohm-cm)}^{-1} \text{ for } La_xSr_{1-x}MnO_3. \qquad (9)$$

This is plotted in Fig. 2 for four values of $x$. We did not have data with which to compare for $La_xSr_xMnO_3$, but Tai, et al.[20] have given such data for a similar compound, $La_xSr_{1-x}Co_{0.2}Fe_{0.8}O_3$, plotted also in Fig. 2. The slightly higher-temperature positions of the experimental peaks would suggest a $W_p$ of 0.19 eV for this compound. The agreement is persuasive support both for the small-polaron description of that compound and for the Emin-Holstein theory. For $La_xSr_{1-x}MnO_3$ these curves might apply up to the Jahn-Teller temperature, where the conductivity rises abruptly as we shall see in the next section. This is just before the $783^\circ$ K at which Eq. (9) reaches its peak, so the conductivity is predicted to increase monatonically with temperature in this range. Zhou and Goodenough[11] measured the conductivity of $LaMnO_3$, without doping, so their increase of over a factor of 100 between $400^oK$ and the Jahn-Teller temperature presumably is dominated by the variation of $N$ with $T$ and does not provide a test. Mahendiran, et al.[17] and Worledge[18] found a conductivity above the Néel temperature which they fit by $\sigma = \sigma_0 exp(-E_a/k_BT)$ rather than the form, Eq. (9). Worledge[18] found an $E_a$ of order 0.09 eV, and Mahendiran found an $E_a$ for $La_{1-x}Ca_xMnO_3$ of 0.16 eV, both above our estimate of $W_p/2$ = 0.07 eV. Worledge's value of $\sigma$ = 40 $(\Omega\text{-cm})^{-1}$ at $300^oK$ is also smaller than the 132 $(\Omega\text{-cm})^{-1}$ we would obtain from Eq. (9) at that temperature with our parameters (and his $x$ = 1/3).

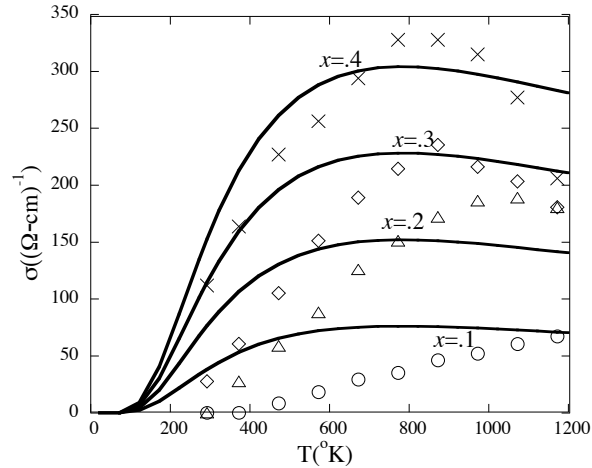

Fig. 2. The lines are plots of Eq. (9). The data are for the similar compound, $La_xSr_{1-x}Co_{0.2}Fe_{0.8}O_3$, From Tai, et al. (Ref. 20).

White[19] has also discussed evidence that in this region and with Worledge's doping of $x$ = 1/3 a superlattice arises with, which may be describable as an ordering of the polarons. He suggests an ordering with the intervening occupied $3z^2-r^2$ orbitals rotated into the $xy$ plane (his Fig. 10.6), which we would not expect because the Jahn-Teller distortion remains, holding them to the $z$ axis. However, the same pattern without rotation would allow an undoing of the Jahn-Teller distortion at each cluster with a doped hole and little disruption of the other cluster sites.

Below the Néel temperature, the observed behavior of the conductivity is totally different. Zhou and Goodenough did not measure below the Néel temperature, but Mahendiran, et al., did[17]. They found that with decreasing temperature the conductivity again rises (there is a peak in resistivity near the Néel temperature[17,18]) and approaches a constant near the absolute zero of temperature. Instead of hopping conductivity, there must be tunneling of holes from cluster to cluster, as in band – or *large polaron* – behavior. This is a formation of bands from the coupled neighboring cluster states as described in the first paragraph of this section. The only difference from band theory, when there are local lattice distortions due to the hole, is a reduction in the coupling, and band width, due to the distortion. . [The role of such distortions in electronic processes is described, for example, in Ref. 15, 309pp.] We noted early in this section that the distortion of each cluster may be thought of as a vibrational mode of the strain coordinate ε and we estimated the frequency, $\omega_0$, of that mode. We now find that when a hole is present in a cluster, the equilibrium ε for that cluster is shifted by a local strain of magnitude $\varepsilon_0 = 0.04$. Then the wavefunction for our system, including the hole, contains also a factor of the harmonic-oscillator wavefunction $\varphi_0(\varepsilon - \varepsilon_0)$ for that cluster, which we take to be the ground-state wavefunction, and a wavefunction $\varphi_0(\varepsilon)$ for every other cluster. Then the coupling between a state of our system with the hole on one site and with the hole on a neighboring site contains factors $S = \int \varphi_0 (\varepsilon - \varepsilon_0) \, \varphi_0 (\varepsilon) d\varepsilon$ for the initial and for the final state, as well as the electronic coupling through the intervening oxygen ion, which we may think of as $V_{eff} = (-V_{pd\sigma}/2)^2/(\varepsilon_d^{maj} - \varepsilon_p)$, but we should do better. We again obtain the tight-binding bands explicitly for these states, based upon the $3z^2 - r^2$ states coupled to σ-oriented $p$ states in the $x$ and $y$ directions by $-V_{pd\sigma}/2$, but each of these Mn-O matrix elements is reduced by a single factor $S$. Then we may expand in wavenumber from the point $k_x = k_y = \pi/2d$ where the bands are maximum to obtain the hole effective mass,

$$m_h / m = \sqrt{2} \hbar^2 / (md^{xy2} \mid SV_{pd\sigma}^{xy} \mid) = 1.53 / S ,$$

$$\text{(10)}$$

neglecting $\varepsilon_d^{maj} - \varepsilon_p$ in comparison to $V_{pd\sigma}$.

To find $S$ we obtain the harmonic-oscillator wavefunctions (as in, for example, Ref. 15, 40ff) as proportional to $\exp(-\varepsilon^2 d^2/2L^2)$, with

$$L^4 = \frac{\hbar^2}{144 M_O c^* d} .$$

$$\text{(11)}$$

$(L/\sqrt{2}$ is the rms zero-point fluctuation, ibid., also a quantity of interest) With the experimental $c^*$ given in the last section and used to estimate $\omega_0$, we find $L = 0.0416$ Å and

$$S = \exp(-\varepsilon_0^2 d^2/4L^2) = 0.393. \tag{12}$$

Substituting this in Eq. (10) gives $m_h/m = 3.89$.

The $1/S$ is a rather large enhancement, not present in LDA-plus-$U$ calculations[5], but could still allow a hole current to flow in the system. If the parabolic-band, effective-mass picture were valid, we would expect the conductivity to increase linearly in doping $x$ but it appears in the experiments[17] to be constant, at moderate $x$. As the temperature rises above 0°K and the spins are not fully aligned, this mass will increase and there will be additional scattering so the conductivity should drop as observed. At some point, seen experimentally to be near the Néel temperature, the hopping conductivity becomes dominant and rises as we have seen in Fig. 2.

At the same time that spin misalignment is reducing in-plane conduction, it is allowing conduction between planes, which we have regarded as zero. Both are also affected by any applied magnetic fields which affect the alignment. This intricate interplay between electrical and magnetic properties is certainly responsible for the colossal magnetoresistance widely studied in these materials. Applying a magnetic field suppresses the antiferromagnetic state, or partially aligns spins above the Néel temperature, and is found to cause large increases in conductivity. The effect persists from low temperatures to far above the Jahn-Teller temperature so it cannot be described by a single mechanism, but is influenced by all of the aspects described here.

## VI. ABOVE THE JAHN-TELLER TRANSITION

Above the Jahn-Teller transition temperature near $750^{\circ}$K, the distortion of the $LaMnO_3$ lattice also disappears, the $3z^2-r^2$ and the $x^2-y^2$ states become equal in energy, leading to half-filled majority-spin $e_g$ bands. We would expect $LaMnO_3$ to become a metallic conductor even without doping. Zhou and Goodenough[11] find that indeed the resistivity drops abruptly at $T_{JT}$ and becomes nearly independent of temperature, as expected for such a metal, but with a resistivity too high to be describable as an ordinary metal. This is not surprising since though there is no ordered Jahn-Teller distortion, we expect very large polaronic distortions associated with the carriers, tying the carrier motion to the lattice distortions. The system might better be described as a polaron liquid than a metal, with fluctuating distortions of the cubic perovskite structure. Experimental studies[21,22] have indicated that the Jahn-Teller state is quite sensitive to the presence of oxygen vacancies, and the state above the Jahn-Teller transition may also be. This would be interesting to explore, but we have not undertaken a treatment of vacancies which would make that possible, nor have we looked further at the properties of such a polaron liquid.

Whatever way one characterizes the conduction properties, we may expect the spins on the individual Mn ions to align in a magnetic field and give paramagnetic susceptibility. Further the same Heisenberg exchange mechanism should apply for the disordered spins. Since we have found the ferromagnetic contribution arising from the interaction between $x^2-y^2$ and $3z^2-r^2$ $e_g$ states to be significantly larger than the antiferromagnetic exchange in $LaMnO_3$, we may expect a ferromagnetic (positive) Curie-Weiss temperature. If in fact we use the weighted average of the $\Delta E_2 = -39.3$ meV which we found for four neighbors in the $xy$ plane with the $+16.4$ eV which we found for the two neighbors in the $z$ direction, we obtain $-20.7$ meV, which with $S=2$ for $LaMnO_3$ in Eq. (7) gives $\theta = 180^{\circ}$K, essentially the same as the ferromagnetic $177^{\circ}$K experimental value given by Zhou and Goodenough.[11] We have no reason to expect such close agreement, but can be gratified that the picture, Fig. 1, for these diverse magnetic and electrical properties holds together so well.

## ACKNOWLEDGEMENT

This work was supported by the Department of Energy under the contract DE-AC26-04NT4187.313.01.05.024, through the Solid State Energy Conversion Alliance.